\documentclass[amssymb,aps]{revtex4}
\usepackage{graphicx}

\begin{document}

\title{Spin-polarized scanning tunneling microscopy of half-metallic ferromagnets:
Non-quasiparticle contributions}
\author{V. Yu. Irkhin}
\affiliation{Institute of Metal Physics, 620219 Ekaterinburg,
Russia}
\author{M. I. Katsnelson}
\affiliation{Institute of Molecules and Materials, University of
Nijmegen, 6525 ED Nijmegen, The Netherlands} 

\begin{abstract}
The role of the many-body (spin-polaronic) effects in the scanning
tunneling spectroscopy of half-metallic ferromagnets (HMF) is
considered. It is shown that the non-quasiparticle (NQP) states
exist in the majority or minority spin gap in the presence of
arbitrary external potential and, in particular, at the surfaces
and interfaces. Energy dependence of the NQP density of states is
obtained in various models of HMF, an important role of the
hybridization nature of the energy gap being demonstrated. The
corresponding temperature dependence of spin polarization  is
calculated. It is shown that the NQP states result in a sharp bias
dependence of the tunneling conductance near zero bias. Asymmetry
of the NQP states with respect to the Fermi energy provides an
opportunity to separate phonon and magnon peaks in the inelastic
spectroscopy by STM.
\end{abstract}

\maketitle
\section{Introduction}

The history of the investigations of half-metallic ferromagnets
(HMF) starts from the electronic structure calculation for NiMnSb
\cite{degroot}; later a number of other examples were discovered,
e.g., CrO$_2$, Fe$_3$O$_4$, a number of the Heusler alloys
Co$_2$MnZ and RMnSb (for a review, see Refs.
\onlinecite{IK,pickett}). These substances have metallic
electronic structure for one spin projection (majority- or
minority-spin states), but for the opposite spin direction the
Fermi level lies in the energy gap. Owing to this fact HMF attract
now a growing attention of researchers in connection with
``spintronics'', or spin-dependent electronics \cite{prinz}. The
spin-up and spin-down contributions to electronic transport
properties have different orders of magnitude, which can result in
a huge magnetoresistance for heterostructures containing HMF
\cite{IK}. Some evidences of the HMF behavior in colossal
magnetoresistance (CMR) materials like La$_{1-x}$Sr$_x$MnO$_3$
found by using tunneling spectroscopy \cite {tun1,tun2} and
photoemission technique \cite{park} have increased considerably
the interest in the half-metallic ferromagnetism; however, the
situation in the CMR systems is controversial, as demonstrate
Andreev reflection experiments \cite{Nag}.

Peculiar band structure of HMF results in an important role of
incoherent (non-quasiparticle, NQP) states which occur because of
correlation effects \cite{IK}. The appearance of NQP states in the
energy gap near the Fermi level \cite{edwards,IK1,AI,IK90} is one
of the most interesting correlation effects typical for HMF. The
origin of these states is connected with ``spin-polaron''
processes: the spin-down low-energy electron excitations, which
are forbidden for HMF in the one-particle picture, turn out to be
possible as superpositions of spin-up electron excitations and
virtual magnons. The density of these states vanishes at the Fermi
level $E_F$ for temperature $T=0$, but increases drastically at
the energy scale of the order of a characteristic magnon frequency
$\overline{\omega}$. The existence of NQP states is relevant for
spin-polarized electron spectroscopy \cite{IK90,AI1}, NMR
\cite{NMR}, core-level spectra of the HMF \cite{core}, and subgap
transport in ferromagnet-superconductor junctions (Andreev
reflection) \cite {falko}. Several experiments could be performed
in order to clarify the impact of the NQP states on spintronics.
In particular, $I-V$ characteristics of half-metallic tunnel
junctions for the case of antiparallel spins are completely
determined by NQP states \cite {ourtransport,falko1}. Recently the
density of NQP states has been calculated from first principles
for a prototype HMF, NiMnSb \cite{lulu}, and for CrAs
\cite{lulu1}.

On the other hand, HMF are very interesting conceptually as a
class of materials which may be convenient to treat many-body
solid state physics that is essentially beyond band theory. It is
accepted that usually many-body effects lead only to
renormalization of the quasiparticle parameters in the sense of
Landau's Fermi liquid (FL) theory, the electronic \textit{liquid}
being \textit{qualitatively} similar to the electron \textit{gas}
(see, e.g., Refs.\onlinecite{Nozieres,VK}). On the other hand, NQP
states in HMF are not described by the FL theory. As an example of
highly unusual properties of the NQP states, we note that they can
contribute to the $T$-linear term in the electron heat capacity
\cite{IK90,IKT}, despite their density at $E_F$ is zero for $T=0$.

Spin-polarized scanning tunneling microscopy (STM) \cite{STM} is a
very efficient new method which enables one to probe directly the
spectral density with spin resolution in magnetic systems. The
spin-polarized STM should be able to probe the NQP states via
their contribution to the differential tunneling conductivity
$d\mathcal{I}/dV$. At zero temperature, NQP states arise only
above $E_F$ for the case of minority-spin gap and only below $E_F$
for the majority-gap HMF \cite{IK}. Unlike the photoemission
spectroscopy which probes only occupied electron states, STM
detects the states both above and below $E_F$, depending on the
sign of bias.

Theoretical investigation of NQP contributions to STM spectra is
the aim of the present paper. The paper is organized as follows.
In Sect.2 we discuss a general expression for the tunneling
current $\mathcal{I}$ as applied to HMF. In Sect.3 the effect of
surface potential and other spatial inhomogeneities on the NQP
spectral density is considered. In Sects.4 and 5 we calculate the
energy and temperature dependences of $d\mathcal{I}/dV$ and treat
the problem of tunneling-current spin-polarization at finite
temperatures. In Sect.6 the bias dependences of the tunneling
conductance are discussed.

\section{Calculation of the tunneling current}

A general expression for the tunneling current in the lowest order
in the tunneling matrix elements has the form \cite{Mah,hamann}
\begin{equation}
\mathcal{I}=\frac{\pi e}{\hbar }\sum\limits_{n\nu \sigma }\left|
M_{n\nu }^{\sigma }\right| ^{2}\int dE\mathcal{A}_{n}^{\sigma
}\left( E\right) \mathcal{A}_{\nu }^{\sigma }\left( E-eV\right)
\left[ f\left( E-eV\right) -f\left( E\right) \right]
\label{curr}
\end{equation}
where $e$ is electron charge, $\sigma $ is the spin projection,
$V$ is the bias, $f\left( \varepsilon \right) $ is the Fermi
distribution function, Greek (Latin) indices label electron
eigenstates for the sample (tip) $\psi _{n\sigma },\psi _{\nu
\sigma }$
\begin{equation}
M_{n\nu }^{\sigma }=\frac{\hbar ^{2}}{2m}\int d\mathbf{A}\left(
\psi _{n\sigma }^{\ast }\nabla \psi _{\nu \sigma }-\psi _{n\sigma
}^{\ast }\nabla \psi _{\nu \sigma }\right)   \label{matelem}
\end{equation}
is the current matrix element, $m$ is the free electron mass (the
surface integral in Eq.(\ref{matelem}) is taken over arbitary area
between the tip and the sample), and
\begin{equation}
\mathcal{A}_{\nu }^{\sigma }\left( E\right) =-\frac{1}{\pi }{\rm
Im}G_{\nu }^{\sigma }\left( E\right)   \label{A}
\end{equation}
is the corresponding spectral density, $G_{\nu }^{\sigma }\left(
E\right) =G_{\nu \nu }^{\sigma }\left( E\right) $ is the sample
Green's function,
\[
G_{\nu \lambda }^{\sigma }\left( E\right) =\langle \langle c_{\nu
\sigma }|c_{\lambda \sigma }^{\dagger }\rangle \rangle_E,
\]
$c_{\nu \sigma }^{\dagger }$ being the creation operators for
conduction electrons. It is worthwhile to emphasize that the
expression (\ref{curr}) takes into account correlation effects in
a formally exact way, assuming that the tunneling probability is
small. In fact, the latter condition should be satisfied for
proper STM measurements, otherwise they cannot be considered as a
true probe. In the WKB approximation Eq.(\ref{curr}) takes the
form \cite{ukr}
\begin{equation}
\mathcal{I}\left( z,V\right) \simeq \frac{\pi e}{\hbar }\left(
\frac{\hbar ^{2}}{2m}\right) ^{2}\sum\limits_{\sigma }\exp \left(
-2z\sqrt{\frac{2m\Phi
_{\sigma }}{\hbar ^{2}}}\right) N_{t}^{\sigma }\left( E_{F}\right) \int dE%
\left[ f\left( E-eV\right) -f\left( E\right) \right] g_{s}^{\sigma
}\left( E\right)   \label{tok}
\end{equation}
where $\Phi_{\sigma} $ is the average of sample and tip work
functions (which is assumed to be large in comparison with $eV$
for simplicity), $z$ is the distance between the surface and the
tip. Here $N_{t}\left( E\right) $ is the density of states (DOS)
of the tip material, which is supposed to be smooth and thus is
replaced by its value at the Fermi energy, and
\begin{equation}
g_{s}^{\sigma }\left( E\right) =\sum\limits_{\nu }\mathcal{A}_{\nu
}^{\sigma
}\left( E\right) \left\langle \nu \sigma \right| \delta \left( \mathbf{k}%
_{\parallel }\right) \left| \nu \sigma \right\rangle   \label{rho}
\end{equation}
is the density of states of the sample with zero in-plane
component of the wave-vector: \ $\mathbf{k}_{\parallel }=0$ so
that the summation is performed only over two points of the Fermi
surface. This condition means that the tunneling probability has a
sharp (at not too small $z$) maximum for the states with velocity
direction normal to the surface. For a generic multi-sheet Fermi
surfaces the condition of the dominant tunneling is, generally
speaking, more complicated, but this modifies only some weakly
bias-dependent factors.

\section{Non-quasiparticle states in inhomogeneous materials}

Since STM probes only surface one has to discuss first
modification of NQP states in comparison with the case of ideal
bulk crystal. The existence of NQP states at the surface of HMF
has been demonstrated for a narrow-band Hubbard model \cite{KE}.
Here we present a general derivation valid in the case of
arbitrary inhomogeneity.

To describe the effects of electron-magnon interaction for the
inhomogeneous case we use the formalism of the exact
eigenfunctions developed earlier for the impurity-state problem in
a ferromagnetic semiconductor \cite{impfm}. The corresponding
Hamiltonian of the $s-d$ exchange model reads
\begin{eqnarray}
\mathcal{H} &=&\int d\mathbf{r}\left( \sum_{\sigma }\Psi _{\sigma
}^{\dagger
}(\mathbf{r})\mathcal{H}_{0}^{\sigma }\Psi _{\sigma }(\mathbf{r}%
)-I\sum_{\sigma \sigma ^{\prime }}\delta \mathbf{S(r)}\Psi
_{\sigma }^{\dagger }(\mathbf{r})\mbox {\boldmath $\sigma
$}_{\sigma \sigma ^{\prime }}\Psi
_{\sigma ^{\prime }}(\mathbf{r})\right) +\mathcal{H}_{d}  \nonumber \\
\mathcal{H}_{0}^{\sigma } &=&-\frac{\hbar ^{2}}{2m}\nabla ^{2}+U_{\sigma }(%
\mathbf{r})  \label{hamilt}
\end{eqnarray}
where $U_{\sigma }(\mathbf{r})$ is the potential energy (with
account of the electron-electron interaction in the mean field
approximation) which is supposed to be spin dependent, $\Psi
_{\sigma }(\mathbf{r})$ is the field operator for the spin
projection $\sigma ,$ $\mathbf{S(r)}$ is the spin density of the
localized-moment system, $\delta \mathbf{S(r)=S(r)-} \left\langle
\mathbf{S(r)}\right\rangle $ is its fluctuating part, the
effect of the average spin polarization\ $\left\langle \mathbf{S(r)}%
\right\rangle $ being included into $U_{\sigma }(\mathbf{r})$. We
use an approximation of contact electron-magnon interaction
described by the $s-d$ exchange parameter $I$,
\begin{equation}
\mathcal{H}_{d}=-\sum_{\mathbf{q}}J_{\mathbf{q}}\mathbf{S}_{\mathbf{q}}%
\mathbf{S}_{-\mathbf{q}}
\end{equation}
is the Heisenberg Hamiltonian of localized spin (for simplicity we
neglect the inhomogeneity effects for the magnon subsystem).

Passing to the representation of the exact
eigenfunctions of the Hamiltonian $\mathcal{H}_{0}^{\sigma },$%
\begin{eqnarray}
\mathcal{H}_{0}^{\sigma }\psi _{v\sigma } &=&\varepsilon _{\nu
\sigma }\psi
_{\nu \sigma },  \nonumber \\
\Psi _{\sigma }(\mathbf{r}) &=&\sum\limits_{\nu }\psi _{\nu \sigma
}\left( \mathbf{r}\right) c_{\nu \sigma },
\end{eqnarray}
one can rewrite the Hamiltonian (\ref{hamilt}) in the following
form:
\begin{equation}
\mathcal{H}=\sum_{\nu \sigma }\varepsilon _{\nu \sigma }c_{\nu
\sigma
}^{\dagger }c_{\Bbb{\nu }\sigma }-I\sum_{\mu \nu \alpha \beta \mathbf{q}%
}\left( \nu \alpha ,\mu \beta |\mathbf{q}\right) \delta
\mathbf{S_{q}}c_{\nu
\alpha }^{\dagger }\mbox {\boldmath $\sigma $}_{\alpha \beta }c_{\mu \beta }+%
\mathcal{H}_{d}
\end{equation}
where
\[
\left( \nu \sigma ,\mu \sigma ^{\prime }|\mathbf{q}\right)
=\left\langle \mu \sigma ^{\prime }\right| e^{i\mathbf{qr}}\left|
\nu \sigma \right\rangle .
\]
We take into account again the electron-spectrum spin splitting in
the mean-field approximation by keeping the dependence of the
eigenfunctions on the spin projection.

We restrict ourselves to the spin-wave region where we can use for
the spin operators the magnon (e.g., Dyson-Maleev) representation.
Then we have for the one-electron Green's function
\begin{equation}
G_{\nu }^{\sigma }(E)=\left[ E-\varepsilon _{\mathbf{\nu }\sigma }-\Sigma _{%
\mathbf{\nu }}^{\sigma }(E)\right] ^{-1}, \label{dys}
\end{equation}
with the self-energy $\Sigma _{\mathbf{\nu }}^{\sigma }(E)$
describing correlation effects.

We start with the perturbation expansion in the electron-magnon
interaction. To second order in $I$ one has
\begin{equation}
\Sigma _{\mathbf{\nu }}^{\sigma }(E)=2I^{2}SQ_{\mathbf{\nu
}}^{\sigma }(E)
\end{equation}
with
\begin{equation}
Q_{\mathbf{\nu }}^{\uparrow }(E)=\sum_{\mu \mathbf{q}}|\left( \nu
\uparrow
,\mu \downarrow |\mathbf{q}\right) |^{2}\frac{N_{\mathbf{q}}+n_{\mathbf{\mu }%
}^{\downarrow }}{E-\varepsilon _{\mathbf{\mu \downarrow }}+\omega _{\mathbf{q%
}}},Q_{\mathbf{\nu }}^{\downarrow }(E)=\sum_{\mu
\mathbf{q}}|\left( \nu
\downarrow ,\mu \uparrow |\mathbf{q}\right) |^{2}\frac{1+N_{\mathbf{q}}-n_{%
\mathbf{\mu }}^{\uparrow }}{E-\varepsilon _{\mu \mathbf{\uparrow }}-\omega _{%
\mathbf{q}}}  \label{q}
\end{equation}
where $n_{\mathbf{\mu }}^{\sigma }=f(\varepsilon _{\mathbf{\mu
}\sigma })$ ,
$\omega _{\mathbf{q}}$ is the magnon energy, $N_{\mathbf{q}}=N_{B}(\omega _{%
\mathbf{q}})$ is the Bose function.

Using the expansion of the Dyson equation (\ref{dys}) we obtain
for the spectral density
\begin{eqnarray}
\mathcal{A}_{\nu \sigma }\left( E\right) &=&-\frac{1}{\pi }\text{\textrm{Im}}%
G_{\mathbf{\nu }}^{\sigma }(E)=\delta (E-\varepsilon _{\mathbf{\nu
}\sigma })
\nonumber \\
&&\ \ \ \ -\delta ^{\prime }(E-\varepsilon _{\mathbf{\nu }\sigma })\text{%
\textrm{Re}}\Sigma _{\mathbf{\nu }}^{\sigma }(E)-\frac{1}{\pi }\frac{\text{%
\textrm{Im}}\Sigma _{\mathbf{\nu }}^{\sigma }(E)}{(E-\varepsilon _{\mathbf{%
\nu }\sigma })^{2}} \label{N(E)}
\end{eqnarray}
The second term in the right-hand side of Eq.(\ref{N(E)}) gives
the shift of quasiparticle energies. The third term, which arises
from the branch cut of the self-energy, describes the incoherent
(non-quasiparticle) contribution owing to scattering by magnons.
One can see that this does not vanish in the energy region,
corresponding to the ``alien'' spin subband with the opposite
projection $-\sigma $.

Neglecting temporarily in Eq.(\ref{q}) the magnon energy $\omega _{\mathbf{q}%
}$ in comparison with typical electron energies and using the
identities
\begin{eqnarray}
\sum_{\mu \mathbf{q}}\frac{|\left( \nu \mu |\mathbf{q}\right) |^{2}}{%
E-\varepsilon _{\mathbf{\mu }}}F\left( \varepsilon _{\mathbf{\mu
}}\right)
&=&-\frac{1}{\pi }\int dE^{\prime }\frac{F\left( E^{\prime }\right) }{%
E-E^{\prime }}\text{\textrm{Im}}\sum_{\mu \mathbf{q}}\frac{|\left( \nu \mu |%
\mathbf{q}\right) |^{2}}{E^{\prime }-\varepsilon _{\mathbf{\mu
}}+i0}
\label{trans} \\
&=&-\frac{1}{\pi }\int dE^{\prime }\frac{F\left( E^{\prime }\right) }{%
E-E^{\prime }}\text{\textrm{Im}}\sum_{\mathbf{q}}\left\langle \nu
\right| e^{i\mathbf{qr}}(E^{\prime
}-\mathcal{H}_{0}+i0)^{-1}e^{-i\mathbf{qr}}\left|
\nu \right\rangle  \nonumber \\
&=&-\frac{1}{\pi }\int dE^{\prime }\frac{F\left( E^{\prime }\right) }{%
E-E^{\prime }}\text{\textrm{Im}}\left\langle \nu \right| (E^{\prime }-%
\mathcal{H}_{0}+i0)^{-1}\left| \nu \right\rangle  \nonumber
\end{eqnarray}
we derive
\begin{eqnarray}
\Sigma _{\nu }^{\uparrow }(E) &=&2I^{2}S\int dE^{\prime
}f(E^{\prime
})\left\langle \nu \uparrow \right| \delta \left( E-E^{\prime }-\mathcal{H}%
_{0}^{\downarrow }\right) \left| \nu \uparrow \right\rangle
\label{sigmaup}
\\
\Sigma _{\nu }^{\downarrow }(E) &=&2I^{2}S\int dE\left[ 1-f(E^{\prime })%
\right] \left\langle \nu \downarrow \right| \delta \left( E-E^{\prime }-%
\mathcal{H}_{0}^{\uparrow }\right) \left| \nu \downarrow
\right\rangle \label{sigmadn}
\end{eqnarray}
Here we restrict ourselves only to the case of zero temperature
$T=0$ neglecting the magnon occupation numbers $N_{\mathbf{q}}$.
Using the tight-binding model for the ideal-crystal Hamiltonian we
find in the real-space representation
\begin{eqnarray}
\Sigma _{\mathbf{R,R}^{\prime }}^{\uparrow }(E) &=&2I^{2}S\int
dE^{\prime
}f(E^{\prime })\left( -\frac{1}{\pi }\text{\textrm{Im}}G_{\mathbf{R,R}%
}^{\downarrow }(E^{\prime })\right) \delta _{\mathbf{R,R}^{\prime
}}
\label{sig_up} \\
\Sigma _{\mathbf{R,R}^{\prime }}^{\downarrow }(E) &=&2I^{2}S\int dE^{\prime }%
\left[ 1-f(E^{\prime })\right] \left( -\frac{1}{\pi }\text{\textrm{Im}}G_{%
\mathbf{R,R}}^{\uparrow }(E^{\prime })\right) \delta
_{\mathbf{R,R}^{\prime }}  \label{sig_dn}
\end{eqnarray}
where $\mathbf{R,R}^{\prime }$ are lattice site indices, and
therefore
\begin{equation}
\Sigma _{\mathbf{\nu }}^{\sigma }(E)=\sum_{\mathbf{R}}\left| \psi
_{\nu \sigma }\left( \mathbf{R}\right) \right| ^{2}\Sigma
_{\mathbf{R,R}}^{\sigma }(E).  \label{sig11}
\end{equation}

Following the method developed by us earlier \cite{IK90,impfm} one
can generalize the above results to the case of arbitrary $s-d$
exchange parameter. Simplifying the sequence of equations of
motion (cf. Ref.\onlinecite{impfm}) we obtain the integral
equation
\begin{equation}
(E-\varepsilon _{\mathbf{\nu }\sigma })G_{\nu \lambda }^{\sigma
}\left( E\right) =\delta _{\nu \lambda }+\sigma IR_{\nu \lambda
}^{\sigma }\left( E\right) -\sigma I\sum_{\kappa }(E-\varepsilon
_{\mathbf{\kappa -}\sigma })R_{\kappa \lambda }^{\sigma }\left(
E\right) G_{\nu \kappa }^{\sigma }\left( E\right) \label{int}
\end{equation}
where
\begin{eqnarray}
R_{\mathbf{\nu \lambda }}^{\uparrow }(E) &=&\sum_{\mu
\mathbf{q}}\left( \mu \downarrow ,\lambda \uparrow
|-\mathbf{q}\right) \left( \nu \uparrow ,\mu
\downarrow |\mathbf{q}\right) \frac{n_{\mathbf{\mu }}^{\downarrow }}{%
E-\varepsilon _{\mathbf{\mu \downarrow }}+\omega _{\mathbf{q}}},
\nonumber
\\
R_{\mathbf{\nu \lambda }}^{\downarrow }(E) &=&\sum_{\mu
\mathbf{q}}\left( \mu \uparrow ,\lambda \downarrow
|\mathbf{q}\right) \left( \nu \downarrow
,\mu \uparrow |\mathbf{q}\right) \frac{1-n_{\mathbf{\mu }}^{\uparrow }}{%
E-\varepsilon _{\mu \mathbf{\uparrow }}-\omega _{\mathbf{q}}}
\label{rr}
\end{eqnarray}
Note that the equation (\ref{int}) is exact in the case of empty
conduction band ($n_{\mathbf{\nu }\sigma }=0,$ one current
carrier, ferromagnetic semiconductor situation), and for finite
band filling this corresponds to a ladder approximation in the
diagram approach.

Similar to (\ref{trans}), we obtain after neglecting the magnon
energies in (\ref{rr}) the  equation for the Green' function
\begin{equation}
\sum_{\kappa }(E-\varepsilon _{\mathbf{\kappa -}\sigma })R_{\kappa
\lambda }^{\sigma }\left( E\right) G_{\nu \kappa }^{\sigma }\left(
E\right)
=\left\langle \nu \sigma \right| R^{\sigma }\left( E\right) (E-\mathcal{H}%
_{0}^{\sigma }+i0)^{-1}G^{\sigma }\left( E\right) \left| \nu
\sigma \right\rangle  \label{ur}
\end{equation}
where we use the matrix notations. Then we have for the operator
Green' function
\begin{equation}
G^{\sigma }\left( E\right) =\left[ E-\mathcal{H}_{0}^{\sigma }+\sigma I(\mathcal{H}%
_{0}^{\sigma }-\mathcal{H}_{0}^{-\sigma })\frac{1}{1+\sigma IR^{\sigma }(E)}%
R^{\sigma }(E)\right] ^{-1}  \label{gm1}
\end{equation}
If we consider spin dependence of electron spectrum in the
simplest rigid-splitting approximation $\varepsilon _{\mathbf{\nu
}\sigma }=\varepsilon _{\mathbf{\nu }}-\sigma I\left\langle
S^{z}\right\rangle $ and thus neglect spin-dependence of the
eigenfunctions $\psi _{\nu \sigma }\left( \mathbf{R}\right) $ the
expressions (\ref{sigmaup}),(\ref{sigmadn}) are drastically
simplified. Then the self-energy does not depend on $\nu $ and we
have
\begin{eqnarray}
\Sigma ^{\sigma }\left( E\right) &=&\frac{2I^{2}SR^{\sigma }(E)}{%
1+\sigma IR^{\sigma }(E)}, \\
R^{\uparrow }(E) &=&\sum_{\mu }\frac{n_{\mathbf{\mu }}^{\downarrow }}{%
E-\varepsilon _{\mathbf{\mu \downarrow }}},\,R^{\downarrow }(E)=\sum_{\mu }%
\frac{1-n_{\mathbf{\mu }}^{\uparrow }}{E-\varepsilon _{\mu \mathbf{\uparrow }%
}}
\end{eqnarray}
If $\mathcal{H}_{0}^{\sigma }$ is just the crystal Hamiltonian
($\nu = \mathbf{k},\varepsilon _{\mathbf{\nu }\sigma
}=t_{\mathbf{k}\sigma}$, $t_{\mathbf{k}\sigma}$ being the band
energy), the expression (\ref{gm1}) coincides with that obtained
in Ref. \onlinecite{IK90} for the Hubbard model after the
replacement $I\rightarrow U$.

The expression (\ref{gm1}) can be also represented in the form
\begin{equation}
G^{\sigma }\left( E\right) =\left[ E-\mathcal{H}_{0}^{-\sigma }-(\mathcal{H}%
_{0}^{\sigma }-\mathcal{H}_{0}^{-\sigma })\frac{1}{1+\sigma
IR^{\sigma }(E)}\right] ^{-1}  \label{gm2}
\end{equation}
The equation (\ref{gm2}) is convenient in the narrow-band case. In
this limit where spin splitting is large
in comparison with the bandwidth of conduction electrons we have $\mathcal{H}%
_{0}^{\uparrow }-\mathcal{H}_{0}^{\uparrow }=-2IS$ and we obtain
for the ``lower'' spin subband with $\sigma =-\mathrm{sign}I$
\begin{equation}
G^{\sigma }\left( E\right) =\left[ E-\mathcal{H}_{0}^{-\sigma }+\frac{2S}{%
R^{\sigma }(E)}\right] ^{-1}
\label{nb}
\end{equation}

For a periodic crystal Eq.(\ref{nb}) takes the form
\begin{equation}
G^{\sigma }_{\bf k}\left( E\right) =\left[ E-t_{{\bf
k}-\sigma}+\frac{2S}{ R^{\sigma }(E)}\right] ^{-1} \label{nb1}
\end{equation}
This expression yields exact result in the limit $I\rightarrow
+\infty $,
\begin{equation}
G^{\downarrow}_{\bf k}\left( E\right) =\left[ \epsilon-t_{\bf
k}+\frac{2S}{R(\epsilon)}\right] ^{-1}, \\
R(\epsilon)=\sum_{\bf k} \frac{1- f(t_{\mathbf
k})}{\epsilon-t_{\mathbf k}} \label{nb22}
\end{equation}
with $ \epsilon=E+IS$, $t_{\mathbf{k}}$ the bare electron
spectrum. In the limit $I\rightarrow -\infty $ Eq.(\ref{nb1})
gives correctly the spectrum of spin-down quasiparticles,
\begin{equation}\label{t*}
G^{\downarrow }_{\bf k}\left( E\right)
=\frac{2S}{2S+1}\left[\epsilon - t^*_{\bf k}\right] ^{-1}
\end{equation}
with $ \epsilon=E-I(S+1),t^*_{\bf k}=[2S/(2S+1)]t_{\bf k}$.
However, it does not describe  the NQP states quite correctly, so
that more accurate expressions can be obtained by using the atomic
representation \cite{orb},
\begin{equation}
G^{\uparrow}_{\bf k}\left( E\right) =\frac{2S}{2S+1}\left[
\epsilon-t^*_{\bf
k}+\frac{2S}{R^*(\epsilon)}\right] ^{-1}, \\
R^*(\epsilon)=\sum_{\bf k} \frac{f(t^*_{\mathbf
k})}{\epsilon-t^*_{\mathbf k}} \label{nb2}
\end{equation}
On the other hand, the result (\ref{nb}), (\ref{nb1}) yields a
good interpolation description in the Hubbard model
\cite{edwards,IK1,IK90}.

The Green's functions (\ref{nb1}), (\ref{nb22}), (\ref{nb2}) have
no poles, at least for small current carrier concentration, and
the whole spectral weight of minority states is provided by the
branch cut (non-quasiparticle states) \cite{IK1,IK90}. For surface
states this result was obtained in Ref.\onlinecite{KE} in a
narrow-band Hubbard model. Now we see that this result can be
derived in an arbitrary inhomogeneous case. For a HMF with the gap
in the minority spin subband NQP states occur above the Fermi
level, and for the gap in the majority spin subband below the
Fermi level.

In the absence of spin dynamics (i.e., neglecting the magnon
frequencies) the NQP density of states has a jump at the Fermi
level. However, the magnon frequencies can be restored in the
final result, in analogy with the case of ideal crystal, which
leads to a smearing of the jump on the energy scale of a
characteristic magnon energy $\overline{\omega }$. It should be
mentioned once more that we restrict ourselves to the case of the
usual three-dimensional magnon spectrum and do not consider the
influence of surface states on the spin-wave subsystem. The
expressions obtained enable us to investigate the energy
dependence of the spectral density.

\section{The non-quasiparticle density of states }

An analysis of the electron-spin coupling yields different
pictures for two possible signs of the $s-d$ exchange parameter
$I$. For $I>0$ the spin-down NQP scattering states form a ``tail''
of the upper spin-down band, which starts from $E_{F}$ (Fig.1)
since the Pauli principle prevents electron scattering into
occupied states.

For $I<0$ spin-up NQP states are present below the Fermi level as
an isolated region (Fig.2): occupied states with the total spin
$S-1$ are a superposition of the states $|S\rangle |\downarrow
\rangle $ and $|S-1\rangle |\uparrow \rangle $. The entanglement
of the states of electron and spin subsystems which is necessary
to form the NQP states is a purely quantum effect formally
disappearing at $S\rightarrow \infty $. To understand better why
the NQP states are formed only below $E_F$ in this case we can
treat the limit $I=-\infty .$ T hen the current carrier is really
a many-body state of the occupied site as a whole with total spin
$S-1/2,$ which propagates in the ferromagnetic medium with spin
$S$ at any site. The fractions of the states $|S\rangle
|\downarrow \rangle $ and $|S-1\rangle|\uparrow \rangle $ in the
current carrier state are $1/(2S+1)$ and $ 2S/(2S+1)$,
respectively, so that the first number is just a spectral weight
of occupied spin-up electron NQP states. At the same time, the
density of empty states is measured by the number of electrons
with a given spin projection which one can add to the system. It
is obvious that one cannot put any spin-up electrons in the
spin-up site at $I=-\infty .$ Therefore the density of NQP states
should vanish above $E_F$.

It is worthwhile to note that in the most of known HMF the gap
exists for minority-spin states \cite{IK}. This is similar to the
case $I>0$, so that the NQP states should arise above the Fermi
energy. For exceptional cases with the majority-spin gap such as a
double perovskite Sr$_{2}$FeMoO$_{6}$ \cite {double} one should
expect formation of the NQP states below the Fermi energy.

It has been proven in the previous section that the presence of
space inhomogeneity (e.g., surface) does not change qualitatively
the spectral density picture, except smooth matrix elements.
Therefore further in this section we will consider, for
simplicity, the case of clean infinite crystal; all the
temperature and energy dependences of the spectral density will be
basically the same for the surface layer.

The second term in the right-hand side of Eq. (\ref{N(E)})
describes the renormalization of quasiparticle energies. The third
term, which arises from the branch cut of the self-energy $\Sigma
_{\nu \sigma }(E)$, describes the incoherent (non-quasiparticle)
contribution owing to scattering by magnons. One can see that this
does not vanish in the energy region, corresponding to the
``alien'' spin subband with the opposite projection $-\sigma $.
Further on we perform for definiteness concrete calculations in
the case  $I>0$ (the case $I<0$ differs, roughly speaking, by a
particle-hole transformation). Summing up Eq.(\ref{N(E)}) to find
the total DOS $N_{\sigma }\left( E\right) $ and neglecting the
quasiparticle shift we obtain
\begin{eqnarray}
N_{\uparrow }(E) &=&\sum_{\mathbf{kq}}\left[ 1-\frac{2I^{2}SN_{\mathbf{q}}}{%
(t_{\mathbf{k+q\downarrow }}-t_{\mathbf{k\uparrow }})^{2}}\right]
\delta
(E-t_{\mathbf{k}\uparrow })  \nonumber \\
N_{\downarrow }(E) &=&2I^{2}S\sum_{\mathbf{kq}}\frac{1+N_{\mathbf{q}}-n_{%
\mathbf{k\uparrow }}}{(t_{\mathbf{k+q\downarrow }}-t_{\mathbf{k\uparrow }%
}-\omega _{\mathbf{q}})^{2}}\delta (E-t_{\mathbf{k}\uparrow }-\omega _{%
\mathbf{q}})  \label{DOS1}
\end{eqnarray}
where we consider for simplicity only second-order perturbation
expression. Basing on a general consideration in the previous
section one can prove that, actually, this expression holds for
arbitrary $I,$ at least, in the framework of $1/2S$ expansion.

The $T^{3/2}$-dependence of the magnon contribution to
the residue of the Green's function, i.e. of the effective
electron mass in the lower spin subband, and an increase with
temperature of the incoherent tail from the upper spin subband
result in a strong temperature dependence of partial densities of
states $N_{\sigma }(E)$, the corrections being of opposite sign.
At the same time, the temperature shift of the band edge for the
quasiparticle states is proportional to $T^{5/2}$ rather than to
magnetization \cite{IK1,impfm}.

The exact solution in the atomic limit (for one conduction
electron), which is valid not only in spin-wave region, but for
arbitrary temperatures, reads \cite{physica83}
\begin{equation}
G^{\sigma }\left( E\right) =\frac{S+1+\sigma \left\langle
S^{z}\right\rangle
}{2S+1}\frac{1}{E+IS}+\frac{S-\sigma \left\langle S^{z}\right\rangle }{2S+1}%
\frac{1}{E-I\left( S+1\right) }.  \label{atom}
\end{equation}
In this case the energy levels are not temperature dependent at
all, whereas the residues are strongly temperature dependent via
the magnetization.

Now we  consider the case $T=0$ for a finite band filling. The
picture of $N(E)$ in HMF (or degenerate ferromagnetic
semiconductor) demonstrates strong energy dependence near the
Fermi level (Figs. 1,2). If we neglect magnon frequencies in the
denominators of Eq.(\ref{DOS1}), the partial density of incoherent
states should occur by a jump above or below the Fermi energy
$E_{F}$ for $I>0$ and $I<0$ respectively owing to the Fermi
distribution functions. An account of finite magnon frequencies
$\omega _{\mathbf{q}}=\mathcal{D}q^{2}$ ($\mathcal{D}$ is the spin
wave stiffness constant) leads to smearing of these singularities,
$N(E_{F})$ being equal to zero. For $|E-E_{F}|\ll \overline{\omega
}$ we obtain

\begin{equation}
\frac{N_{-\alpha }(E)}{N_\alpha (E)}=\frac 1{2S}\left| \frac{E-E_F}{%
\overline{\omega }}\right| ^{3/2}\theta (\alpha (E-E_F)),\,\alpha =\text{%
\textrm{sgn}}I  \label{alpha1}
\end{equation}
($\alpha =\pm $ corresponds to the spin projections $\uparrow ,\downarrow $%
). With increasing $|E-E_F|,\,N_{-\alpha }/N_\alpha $ tends to a
constant value which is of order of $I^2$ within the perturbation
theory.

In the strong coupling limit where $|I|\rightarrow \infty $ we
have from (\ref{DOS1})
\begin{equation}
\frac{N_{-\alpha }(E)}{N_{\alpha }(E)}=\frac{1}{2S}\theta (\alpha
(E-E_{F})),|E-E_{F}|\gg \overline{\omega }  \label{alpha2}
\end{equation}
In fact, this expression is valid only in the framework of the
$1/2S$-expansion, and in the narrow-band quantum case we have to
use more exact expressions (\ref{nb22}),(\ref {nb2}). In numerical
calculations, we follow to Ref.\onlinecite{orb} and smear the
resolvents,
\[
R(E)\rightarrow \overline{R}(E)=\int d\omega K(\omega )R(E\pm
\omega)
\]
We use the semielliptic magnon DOS $K(\omega )$ which is
proportional (with the corresponding shift) to the bare electron
DOS, the maximum magnon frequency being determined by the electron
concentration $c$ \cite{orb}. This approximation provides the
correct behavior near the Fermi level (cf. Ref.\onlinecite{iz04}),
although gives an unphysical shift of the band bottom by the
maximum magnon frequency.

The results of calculations are shown in Figs.3, 4. One can see
that in the model with $I\rightarrow-\infty$ (for $S=1/2$ this is
equivalent to the Hubbard model with the replacement $t_{\bf k}
\rightarrow t_{\bf k}/2$, see Refs.\onlinecite{IK1,IK90}) the
``Kondo" peaks \cite{iz04} modify considerably the picture. Note
that the function $-(1/\pi) \rm{Im} \overline{R^*}(E)$, which
yields DOS in the lowest-order approximation in the electron
concentration, does not have such peaks.

To investigate  details of the energy dependence of $N\left(
E\right) $ in the broad-band case  we assume the simplest
isotropic approximation for the majority-spin electrons,
\begin{equation}
t_{\mathbf{k}\uparrow }-E_{F}\equiv \xi _{\mathbf{k}}=\frac{k^{2}-k_{F}^{2}}{%
2m^*}.  \label{DOSbareup}
\end{equation}
Provided  that we use the rigid splitting approximation
$t_{\mathbf{k}\downarrow }=t_{ \mathbf{k}\uparrow }+\Delta $
($\Delta =2IS,I>0$), the half-metallic situation (or, more
precisely, the situation of degenerate ferromagnetic
semiconductor) takes place for $\Delta >E_{F}$. Then qualitatively
the equation (\ref{alpha1}) works to accuracy of a prefactor. It
is worthwhile to note that, formally speaking, the NQP
contribution to DOS occurs also for an ``usual" metal where
$\Delta <E_{F}$. In the case of small $\Delta $ there is a
crossover energy (or temperature) scale
\begin{equation}
T^{\ast }=\mathcal{D}\left(m^* \Delta/ k_F\right) ^{2}
\label{crossover}
\end{equation}
which is the magnon energy at the boundary of Stoner continuum,
$T^{\ast
}\simeq \overline{\omega }\left( \Delta /E_{F}\right) ^{2}\ll \overline{%
\omega }.$ At $|E-E_{F}|\ll \overline{\omega }$ the
equation (\ref {DOS1}) for the NQP contribution reads
\begin{equation}
\delta N_{\downarrow }(E)\propto \left[ \frac{1}{2}\ln \left| \frac{1+\sqrt{%
\left( E-E_{F}\right) /T^{\ast }}}{1-\sqrt{\left( E-E_{F}\right) /T^{\ast }}}%
\right| -\sqrt{\left( E-E_{F}\right) /T^{\ast }}\right] \theta
(E-E_{F}). \label{DOS111}
\end{equation}
At $|E-E_{F}|\ll T^{\ast }$ this gives the same results as above.
However, at $T^{\ast }$ $\ll |E-E_{F}|\ll \overline{\omega }$ this
contribution is proportional to $-\sqrt{\left( E-E_{F}\right)
/T^{\ast }}$ and is \textit{negative} (of course, the total DOS is
always positive). This demonstrates that one should be very
careful when discussing the NQP states for the systems which are
not half-metallic.

The model of rigid spin splitting used above is in fact not
applicable for the real HMF where the gap has a hybridization
origin \cite{degroot,IK}. The simplest model for HMF is as
follows: a ``normal'' metallic spectrum for majority electrons
(\ref{DOSbareup}) and the hybridization gap for minority ones,
\begin{equation}
t_{\mathbf{k}\downarrow }-E_{F}=\frac{1}{2}\left( \xi _{\mathbf{k}}+\mathrm{%
sgn}\left( \xi _{\mathbf{k}}\right) \sqrt{\xi _{\mathbf{k}}^{2}+\Delta ^{2}}%
\right)   \label{DOSbaredn}
\end{equation}
Here we assume for simplicity that the Fermi energy lies exactly
in the middle of the hybridization gap (otherwise one needs to
shift $\xi _{\mathbf{k}}\rightarrow \xi _{\mathbf{k}}+E_{0}-E_{F}$
in the last equation, $E_{0}$ being the middle of the gap). One
can replace in Eq.(\ref{DOS1})  $\xi _{\mathbf{k+q}}$ by
$\mathbf{v}_{\mathbf{k}}\mathbf{q}$,
$\mathbf{v}_{\mathbf{k}}=\mathbf{k}/m^*$. First, we integrate over
the angle between the vectors $\mathbf{k}$ and $\mathbf{q.}$ It is
easy to calculate
\begin{equation}
\left\langle \left( \frac{1}{t_{\mathbf{k+q\downarrow }}-t_{\mathbf{%
k\uparrow }}-\omega _{\mathbf{q}}}\right) ^{2}\right\rangle =\frac{8}{%
v_{F}q\Delta }\left( \frac{2}{3}\left[
X^{3}-(X^{2}+1)^{3/2}+1\right] +X\right)   \label{angle}
\end{equation}
where angular brackets stand for the average over the angles of
the vector $\mathbf{k}$, $X=k_{F}q/m^*\Delta .$ Here we do have
the crossover with the energy scale $T^{\ast }$ which can be small
for small enough hybridization gap. For example, in NiMnSb the
conduction band width is about 5 eV and the distance from the
Fermi level to the nearest gap edge (i.e. indirect energy gap
which is proportional to $\Delta^2$) is smaller than 0.5 eV, so
that $(\Delta /E_{F})^2\leq 0.1$.

For the case $0<E-E_{F}\ll \overline{\omega }$ one has
\begin{equation}
N_{\downarrow }(E)\propto b\left(\frac{ E-E_{F}} {T^{\ast
}}\right),
\\
b(y)= \frac{2}{5}\left[ y^{5/2}-\left( 1+y\right) ^{5/2}+1\right]
+y+y^{3/2}=\left\{
\begin{array}{cc}
y^{3/2}, & y\ll 1 \\
y, & y\gg 1
\end{array}
\right.
\label{ss}
\end{equation}
The function $b(x)$ is shown in Fig.5. Thus the behavior
$N_{\downarrow }(E)\propto \left( E-E_{F}\right) ^{3/2}$ takes
place only for very small excitation energies $E-E_{F}\ll T^{\ast
}$, whereas in a broad interval $T^{\ast }$ $\ll E-E_{F}\ll
\overline{\omega }$ one has the linear dependence $N_{\downarrow
}(E)\propto E-E_{F}.$

\section{The temperature dependence of  spin polarization}

Simple qualitative considerations \cite{edw}, as well as direct
Green's functions calculations \cite{AI,aus1} for magnetic
semiconductors, demonstrate that spin polarization of conduction
electrons in the spin-wave region is proportional to magnetization
\begin{equation}
P\equiv \frac{N_{\uparrow }-N_{\downarrow }}{N_{\uparrow }+N_{\downarrow }}%
=2P_0\langle S^z\rangle  \label{polar1}
\end{equation}
A weak ground-state depolarization $1-P_0$ occurs in the case
where $I>0$. The behavior $P(T)\simeq \langle S^z\rangle $ is
qualitatively confirmed by experimental data on field emission
from ferromagnetic semiconductors \cite {kisker} and transport
properties of half-metallic Heusler alloys \cite{otto}.

An attempt was used \cite{skomski} to generalize the result
(\ref{polar1}) on the HMF case (in fact, using qualitative
arguments which are valid only in the atomic limit, see
Eq.(\ref{atom})). However, we will demonstrate that the situation
for HMF is more complicated.

In this section we focus on the magnon contribution to DOS
(\ref{DOS1}) and calculate the function
\begin{equation}
\Lambda =\sum_{\mathbf{kq}}\frac{2I^{2}SN_{\mathbf{q}}}{(t_{\mathbf{%
k+q\downarrow }}-t_{\mathbf{k\uparrow }}-\omega
_{\mathbf{q}})^{2}}\delta (E_{F}-t_{\mathbf{k}\uparrow })
\label{phi}
\end{equation}
Using the parabolic electron spectrum $t_{\mathbf{k\uparrow
}}=k^{2}/2m^*$ and averaging over the angles of the vector
$\mathbf{k}$ we obtain
\begin{equation}
\Lambda =\frac{2I^{2}Sm^{2}}{k_{F}^{2}}\rho \sum_{\mathbf{q}}\frac{N_{\mathbf{q}%
}}{\left( q^{\ast }\right) ^{2}-q^{2}}  \label{phi1}
\end{equation}
where $\rho =N_{\uparrow }(E_{F},T=0),$ we have used the condition
$q\ll k_{F},\,q^{\ast }=m^*\Delta /k_{F}=\Delta /v_{F},$ where
$\Delta =2\left| I\right| S$ is the spin splitting$.$ In the
ferromagnetic semiconductor we have, in agreement with the
qualitative considerations presented above:
\begin{equation}
\Lambda =\frac{S-\langle S^{z}\rangle }{2S}\rho \propto \left( \frac{T}{T_{C}}%
\right) ^{3/2}\rho   \label{phi3}
\end{equation}

Further on we consider the spectrum model (\ref{DOSbareup}), (\ref
{DOSbaredn}) where the gap has a hybridization origin.  At $T\ll
T^{\ast }$ we reproduce the result (\ref{phi3}) which is actually
universal for this temperature region. \ At $T^{\ast }$ $\ll T\ll
\overline{\omega }$ we derive
\begin{equation}
\Lambda =\sum_{\mathbf{kq}}2I^{2}SN_{\mathbf{q}}\delta (\xi _{\mathbf{k}})\frac{%
16}{3v_{F}q\Delta }\propto q^{\ast }\sum_{\mathbf{q}}\frac{N_{\mathbf{q}}}{q}%
\propto \frac{T^{\ast 1/2}}{T_{C}^{1/2}}T\ln \frac{T}{T^{\ast }}
\label{phi11}
\end{equation}
This result distinguishes  HMF like the Heusler alloys from
ferromagnetic semiconductors and narrow-band saturated
ferromagnets. In the narrow-band case the spin polarization
follows the magnetization up to the Curie temperature $T_{C}$.

For finite temperatures the density of NQP states at the Fermi
energy  is proportional to \cite{edw,AI,IKT}
\begin{equation}
N(E_F) \propto \int_0^\infty d\omega \frac{K(\omega)}{\sinh
(\omega /T)}  \label{atFermi}
\end{equation}
Generally, for temperatures which are comparable with the Curie
temperature $T_{C}$ there are no essential difference between
half-metallic and ``ordinary'' ferromagnets since the gap is
filled. The corresponding symmetry analysis is performed in Ref.
\onlinecite{IKT} for a model of conduction electrons interacting
with ``pseudospin'' excitations in ferroelectric semiconductors.
The symmetrical (with respect to $E_{F}$) part of $N(E)$ in the
gap can be attributed to smearing of electron states by
electron-magnon scattering; the asymmetrical (``Kondo-like") one
is the density of NQP states owing to the Fermi distribution
function. Note that this filling of the gap is very important for
possible applications of HMF in spintronics: they really have some
advantages only provided that $T\ll T_{C}$. Since a
single-particle Stoner-like theory leads to much less restrictive
(but unfortunately completely wrong) inequality $T\ll \Delta ,$
the many-body treatment of the spin-polarization problem
(inclusion of collective spin-wave excitations) is required.

\section{Bias dependence of the tunneling conductance}

Now we consider an application of  the results obtained above to
the tunneling spectroscopy problem. The formulas of Sect.4 for the
energy dependence of NQP contributions are, strictly speaking,
derived for the usual one-electron density of states at $E_F$,
which is observed, say, in photoemission measurements. However,
the factor of $g_{s}^{\sigma }\left( E\right)$, which is present
in the expression for the tunneling current (\ref{tok}), does not
influence the temperature dependence, and therefore these results
are valid for spin polarization from tunneling conductance at zero
bias in STM.

The only difference in the NQP contributions to $g_{s}^{\sigma
}\left( E\right) $ and $N_{\sigma }\left( E\right) $ is in that
after summation over the magnon wavevector $\mathbf{q}$ the
integration is performed over not in the the whole Fermi surface,
but its two points (see Eq.(\ref{rho})). For a spherical Fermi
surface for majority electrons the results differ by the constant
factor of the Fermi surface diameter. However, the energy and
temperature dependences should be the same in a more general case.

Consider the bias dependence of the tunneling current for
zero temperature. One can see from Eq.(\ref{tok}) that
\begin{equation}
\frac{d\mathcal{I}^{\sigma }\left( V\right) }{dV}\propto
g_{s}^{\sigma }\left( eV\right) \propto N_{\sigma }\left(
eV\right)   \label{prop}
\end{equation}
Again, the last proportionality can be strictly justified in the
case of a spherical Fermi surface only, but is qualitatively valid
for arbitrary electron spectrum.

One should keep in mind that sometimes the surface of HMF is not
half-metallic; in particular, this is the case of a prototype HMF,
NiMnSb \cite{surface}. In such a situation, the tunneling current
for minority electrons is due to the surface states only. However,
the NQP states can be still visible in the tunneling current via
the hybridization of the bulk states with the surface one. The
hybridization lead to the Fano antiresonance picture which is
usually observed in STM investigations of the Kondo effect at
metallic surfaces (see, e.g., Refs. \onlinecite{zaw,mad,kol}). In
these cases the tunneling conductance will be proportional to a
mixture of $N_{\sigma }\left( eV\right) $ and $L_{\sigma }\left(
eV\right)$, $L_{\sigma }\left( E\right) $ being the real part of
the on-site Green's function,
\begin{equation}
L_{\sigma }\left( E\right) =\mathcal{P}\int dE^{\prime
}\frac{N_{\sigma }\left( E^{\prime }\right) }{E-E^{\prime }}.
\label{real}
\end{equation}
($\mathcal{P}$ stands for principal value, $E$ is referred to the
Fermi energy). Surprisingly, in this case the effect of NQP states
on the tunneling current can be even more pronounced in comparison
with the ideal crystal. The reason is that the analytical
continuation of the jump in $N_{\sigma }\left( E\right) $ is
logarithm; both singularities are cut at the energy
$\overline{\omega };$ nevertheless, the energy dependence of
$L_{\sigma }\left( E\right) $ can be pronounced, see Fig.6. This
is similar to the effect of enhancement of the NQP contribution to
the x-ray absorption and emission spectra, which was predicted in
Ref. \onlinecite{core}.

Now we discuss in more detail the energy dependence for $|E|\ll \overline{%
\omega }$. The analytical continuation  of the $E^{3/2}\theta (E)$%
-contribution to $N_{\sigma }\left( E\right) $ yields the contribution  $%
(-E)^{3/2}\theta (-E)$ in $L_{\sigma }\left( E\right) $ which is
non-zero on the other side with respect to $E_F$ (a situation that
is formally similar to the electronic topological transition, see
Ref.\onlinecite{tref}). The one-sided linear dependence in
$N_{\sigma }\left( E\right) $ according to Eq.(\ref{ss})
corresponds to $E\ln |E|$ in $L_{\sigma }\left( E\right)$.

STM measurements of electron DOS give also an opportunity to probe
\textit{bosonic} excitations interacting with the conduction
electrons. Due to electron-phonon coupling, the derivative
$dN_{\sigma }\left( E\right) /dE$ and thus
$d^{2}\mathcal{I}^{\sigma }\left( V\right) /dV^{2}$ at $eV=E$ have
peaks at the energies $E=\pm \omega _{i}$ corresponding to the
peaks in the phonon DOS. According to our results (see, e.g.,
Eq.(\ref{DOS1}), the same effect should be observable for the case
of electron-magnon interaction. However, in the latter case these
peaks are essentially asymmetric with respect to the Fermi energy
(zero bias) due to asymmetry of the non-quasiparticle
contributions. This asymmetry can be used to distinguish phonon
and magnon peaks.

\section{Conclusions}

In the present paper we have demonstrated that non-quasiparticle
states in half-metallic ferromangnets exist not only for an ideal
crystal, but also in the presence of an arbitrary external
potential. In particular, they occur at the surface of the
half-metallic ferromagnets. These states can be probed by the STM
both directly and via their effect on the surface states (the Fano
antiresonance case). Therefore, they can be observable even for
the situation of surface ``dead layers'' where the surface is not
half-metallic. The expressions obtained can be used for realistic
electronic structure calculations of NQP contributions to the
electron energy spectrum of the surfaces of HMF.

Temperature dependence of the spin polarization at the Fermi
energy which can be also probed by the STM follows the temperature
dependence of magnetization an very low temperatures. For the HMF
with a hybridization gap, there is a crossover energy
(temperature) $T^{\ast }\ll T_{C}$ where the character of the
temperature dependence is changed. The energy dependence of the
NQP contributions (and consequently the bias dependence of the
tunneling current) is  strongly influenced by the band structure
too. In particular, for HMF with a hybridization gap this
demonstrates a linear rather than $E^{3/2}$ behavior in a wide
interval. In the narrow-band case a Kondo-like peak (Fig.4) near
the Fermi level should be observed in tunneling experiments.

Due to asymmetry of NQP states with respect to the Fermi
energy, the magnon peaks in $d^{2}\mathcal{I}^{\sigma }\left(
V\right) /dV^{2}$ are also asymmetric  with respect to the zero
bias, in contrast with the phonon ones. This gives an opportunity
to distinguish between phonon and magnon peaks in the inelastic
spectroscopy by STM.

In principle, the NQP effects discussed should exist also in usual
metallic ferromagnets. However, only in HMF they can be picked up
in a pure form.

The research described was supported in part by Grant
No.~747.2003.2 (Support of Scientific Schools) from the Russian
Basic Research Foundation and by the Netherlands Organization for
Scientific Research (Grant NWO 047.016.005).

\newpage

\begin{figure}[tbp]
\includegraphics[clip]{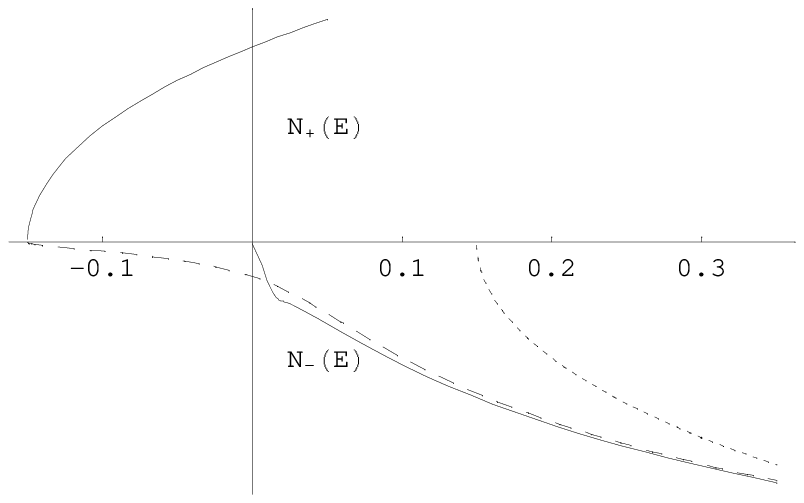}
\caption{ Density of states in the $s-d$ exchange model of a
half-metallic ferromagnet with $S=1/2,I=0.3$ for the semielliptic
bare band with the width of $W=2$. The  Fermi energy calculated
from the band bottom is 0.15 (the energy is referred to $E_F$).
The magnon band is also assumed semielliptic with the width of
$\omega_{\rm max}=0.02$. The non-quasiparticle tail of the
spin-down subband (lower half of the figure) occurs above the
Fermi level. The corresponding picture for the empty conduction
band is shown by dashed line; the short-dashed line corresponds to
the mean-field approximation.} \label{fig:1}
\end{figure}

\begin{figure}[tbp]
\includegraphics[clip]{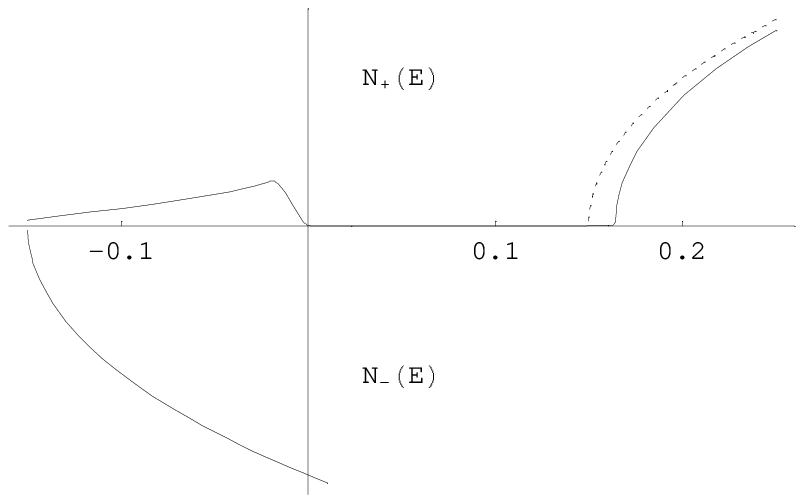}
\caption{Density of states in a half-metallic ferromagnet with
$I=-0.3<0$, other parameters being the same as in Fig.1.  The
spin-down subband (lower half of the figure) nearly coincides with
the bare band shifted by $IS$. Non-quasiparticle states in the
spin-up subbands (upper half of the figure) occur below the Fermi
level; the short-dashed line corresponds to the mean-field
approximation.} \label{fig:2}
\end{figure}

\begin{figure}[tbp]
\includegraphics[clip]{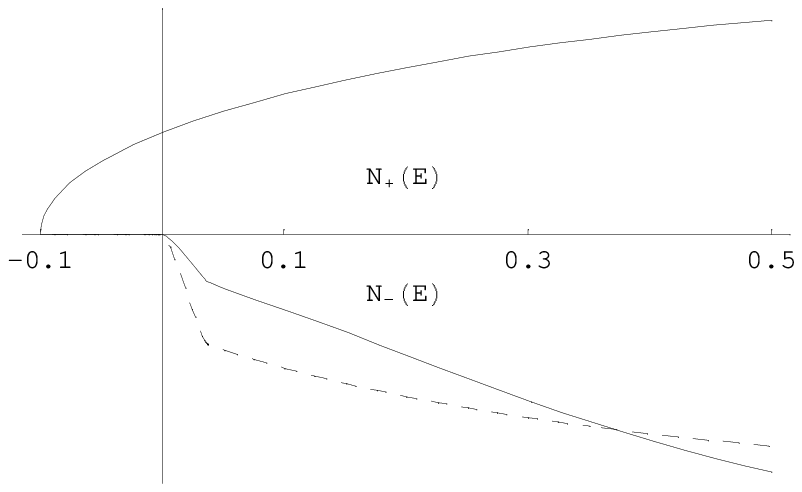}
\caption{Density of states in a half-metallic ferromagnet in the
s-d model with $I\rightarrow+\infty, S=1/2$. The Fermi energy
calculated from the bare band bottom is 0.1 (concentration of
conduction electrons is $c=0.019$).  The dashed line shows the
function $-(1/\pi) \rm{Im}\overline{R}(E)$.} \label{fig:3}
\end{figure}

\begin{figure}[tbp]
\includegraphics[clip]{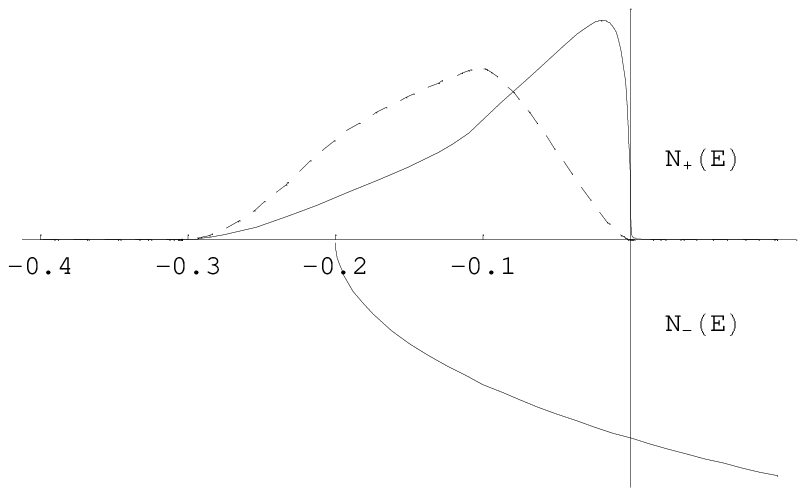}
\caption{Density of states in a half-metallic ferromagnet in the
s-d model with $I\rightarrow-\infty, S=1/2$.The Fermi energy
calculated from the bare band bottom is 0.2 ($c=0.034$). The
dashed line shows the function $-(1/\pi) \rm{Im}
\overline{R^*}(E)$. } \label{fig:4}
\end{figure}

\begin{figure}[tbp]
\includegraphics[clip]{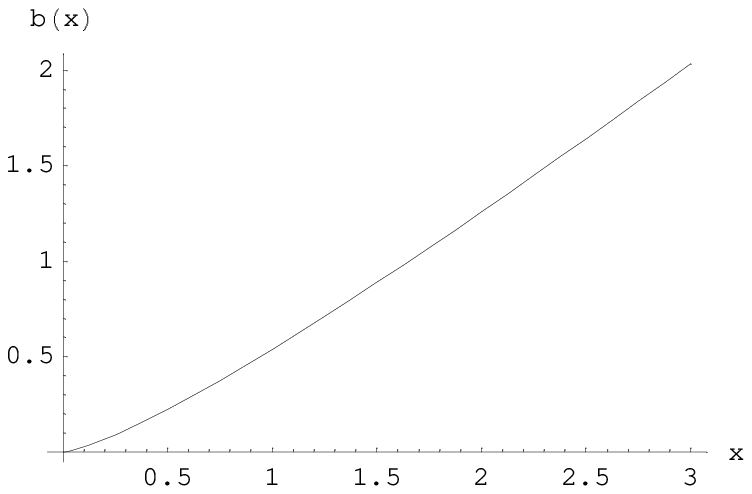}
\caption{Plot of the function $b(x)$. } \label{fig:5}
\end{figure}

\begin{figure}[tbp]
\includegraphics[clip]{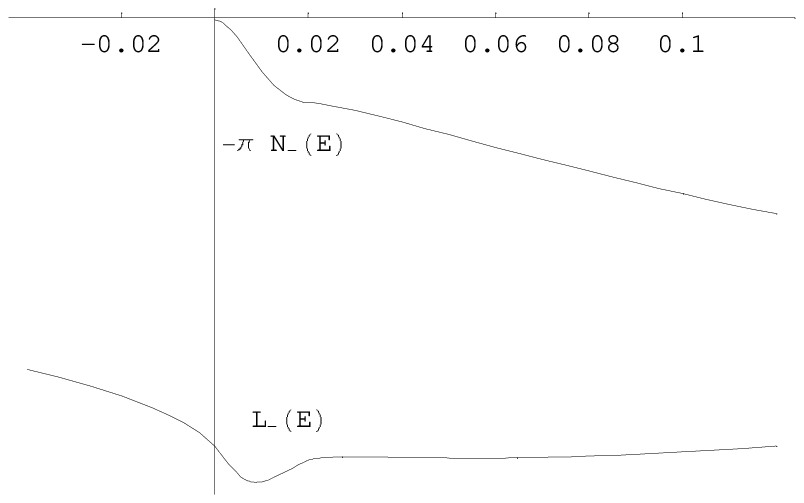}
\caption{ Plot of the imaginary  (upper line) and real
 (lower line) parts of the Green's function near the Fermi level in a
half-metallic ferromagnet with the same parameters as in Fig.1. }
\label{fig:6}
\end{figure}

\end{document}